\documentclass[US Letter]{article}
\usepackage{spconf,amsmath,graphicx,amssymb,delarray,subfigure,verbatim,booktabs}
\usepackage{diagbox, makecell}
\usepackage[numbers,compress]{natbib}
\usepackage{multirow}
\usepackage{bbding}
\usepackage[english]{babel}
\usepackage{lipsum}

\title{Embedding and Beamforming: All-neural Causal Beamformer for Multichannel Speech Enhancement}
\name{Andong Li$^{\star \dagger}$, Wenzhe Liu$^{\star \dagger}$, Chengshi Zheng$^{\star \dagger}$\thanks{Chengshi Zheng is the corresponding author.}, Xiaodong Li$^{\star \dagger}$}
\address{$^{\star}$ Key Laboratory of Noise and Vibration Research, Institute of Acoustics, Chinese Academy\\
	of Sciences, Beijing, China\\
	$^{\dagger}$ University of Chinese Academy of Sciences, Beijing, China}
\begin{document}
\ninept
\maketitle
\newcommand\blfootnote[1]{%
	\begingroup
	\renewcommand\thefootnote{}\footnote{#1}%
	\addtocounter{footnote}{-1}%
	\endgroup
}

\begin{abstract}
The spatial covariance matrix has been considered to be significant for beamformers. Standing upon the intersection of traditional beamformers and deep neural networks, we propose a causal neural beamformer paradigm called \emph{Embedding and Beamforming}, and two core modules are designed accordingly, namely EM and BM. For EM, instead of estimating spatial covariance matrix explicitly, the 3-D embedding tensor is learned with the network,  where both spectral and spatial discriminative information can be represented. For BM, a network is directly leveraged to derive the beamforming weights so as to implement filter-and-sum operation. To further improve the speech quality, a post-processing module is introduced to further suppress the residual noise. Based on the DNS-Challenge dataset, we conduct the experiments for multichannel speech enhancement and the results show that the proposed system outperforms previous advanced baselines by a large margin in multiple evaluation metrics.
\end{abstract}
\vspace{-0.3cm}
\section{Introduction}
\label{sec:intro}
\vspace{-0.3cm}
Speech enhancement (SE) attempts to extract the target speech from the mixture signals. Due to the utilization of spatial information to distinguish between target and interference, a plethora of beamforming-based multichannel speech enhancement algorithms have been widely proposed in a diverse set of applications, ranging from audio-video conferencing to human-machine interaction~{\cite{gannot2017consolidated, van1988beamforming}}.

With the renaissance of deep neural networks (DNNs), neural beamformers have propitiated wide interest due to their promising performance in speech restoration and automatic speech recognition (ASR) systems~{\cite{heymann2016neural, wang2020complex, hori2017multi}}. A typical strategy is to combine DNNs with traditional beamforming techniques. Specifically, a single-channel SE network is first adopted to parallelly estimate time-frequency (T-F) masks \emph{w.r.t.} speech and noise for each channel. The spatial covariance matrices are then calculated to obtain the optimal weights for beamformers based on statistical optimization criteria~{\cite{heymann2016neural, erdogan2016improved, heymann2015blstm}}, like minimum variance distortionless response (MVDR) beamformers and multichannel wiener filter (MWF) beamformers. However, as the second stage is purely based on statistical theory and is usually irrelevant to the mask estimation, the pre-estimation error may heavily hamper the subsequent beamforming results. More recently, regression-based approaches began to thrive. Instead of obtaining the beamformers' weights, the spatial information is represented either manually or implicitly, which serves as the auxiliary aspect of spectral feature to facilitate the speech recovery in either the time domain~{\cite{zhang2020end, gu2019neural}} or T-F domain~{\cite{wang2018combining, fu2021desnet}}. The overall network topology is akin to the single-channel case. Nonetheless, it is still far from affirmative how to combine the spatial and spectral features efficiently~{\cite{gu2019neural}}. Moreover, the potential of spatial filtering is not fully exploited, which limits the overall enhancement performance in real acoustic scenarios.

Recently, another research line follows the guidance of estimating the beamforming weights with DNNs. As an early trial, Xiao~\emph{et al.}~\cite{xiao2016deep} proposed to use GCC features to estimate the weights with DNNs. In~{\cite{luo2019fasnet}}, Luo~\emph{et al.} proposed the FasNet to implement the filter-and-sum operation in the time domain. Nonetheless, it lacks adequate robustness and superiority compared with advanced T-F domain based works~{\cite{fu2021desnet}}. More recently, an all-deep-learning beamforming paradigm called ADL-MVDR was proposed, where the mask estimation, spatial covariance calculation, and framewise beamforming are integrated into a whole network and can be trained end-to-end~{\cite{zhang2021adl}}. Despite the impressive performance in speech quality and ASR accuracy, as only the supervision \emph{w.r.t.} target speech is provided, it remains agnostic whether the internal signal-theory based operations follow the expected physical definitions.  

At this point, we would like to answer a question, \emph{i.e.,} \emph{how to guarantee a neural system that can generate frame-level weights for beamforming?} We argue that it should meet two requirements. First, it should incorporate abundant spatial information to distinguish the sources from different directions. Besides, as the filters need to be updated at each frame, it should learn the T-F cue to assist the separation between speech and interference especially when the spatial cue is absent and blurred. To this end, we propose a generalized framework with the causal setting called \textbf{E}mbedding \textbf{a}nd \textbf{B}eamforming Network (\textbf{EaBNet}) for all-neural beamforming. Two core modules are designed, namely \emph{Embedding Module} (EM) and \emph{Beamforming Module} (BM). In the first module, the network aims to extract the feature from the spectral and spatial perspectives and obtain the 3-D embedding tensor that can latently distinguish between speech and noise components in both senses. In the second module, rather than obtain the filter weights following the statistically optimal beamformer theory, we adopt a network to accomplish the process. It has been illustrated that DNNs can better learn the optimal filter weights than following the traditional beamformer formulas~{\cite{xu2021generalized}}. Note that in contrast to~{\cite{xu2021generalized}} that the second-order spatial statistics need to be explicitly calculated, our approach chooses to directly obtain the temporal-spatial embedding so that it can potentially learn higher-order spatial statistics with data-driven. With the experiments on DNS-Challenge corpus, our system outperforms previous state-of-the-art (SOTA) baselines by a large margin and also surpasses the MVDR beamformer with oracle masks. 

The rest of the paper is organized as follows. In Section~{\ref{sec:physical-model}}, we formulate the physical model. In Section~{\ref{sec:proposed-system}}, the proposed system is introduced in detail. Section~{\ref{sec:experimental-setup}} gives the experimental setup, and the experimental results and analysis are provided in Section~{\ref{sec:results-and-analysis}}. Some conclusions are drawn in Section~{\ref{sec:conclusion}}.
\begin{figure*}[t]
	\centering
	\centerline{\includegraphics[width=1.50\columnwidth]{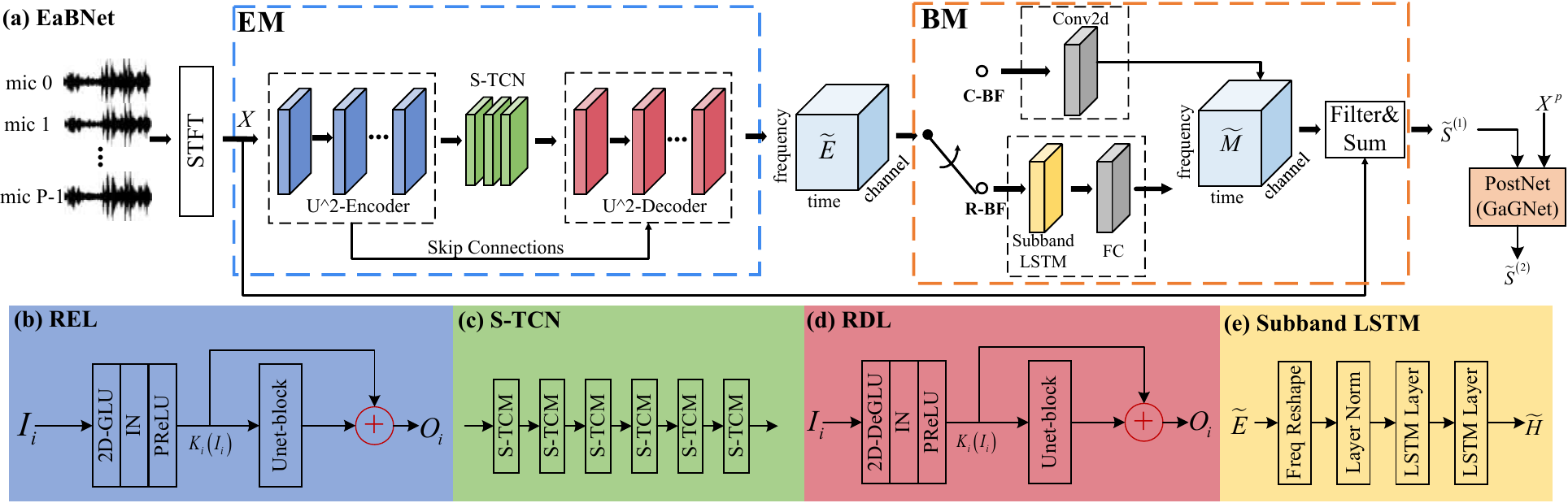}}
	\caption{The diagram of the proposed framwork EaBNet. It mainly consists of two modules, namely EM and BM. Besides, the post-processing module is also introduced to further suppress the residual noise component. Different modules are remarked with different colors.}
	\label{fig:architecture}
	\vspace{-0.6cm}
\end{figure*}
\vspace{-0.42cm}
\section{PHYSICAL MODEL}
\vspace{-0.3cm}
\label{sec:physical-model}
Let us assume $x^{p}(t)$, with $p = 0,\cdots,P-1$, denotes the time-domain noisy and reverberant speech signal at the $p$th microphone. The physical model in the short-time Fourier transform (STFT) domain can be given by:
\vspace{-0.2cm}
\begin{equation}
\label{eqn1}
\mathbf{X}_{f,t} = \mathbf{S}_{f,t} + \mathbf{N}_{f,t} = \mathbf{c}_{f}S_{f,t} + \mathbf{r}_{f}N_{f,t},
\vspace{-0.2cm}
\end{equation}
\vspace{-0.0cm}
where $\left\{ \mathbf{X}_{f,t}, \mathbf{S}_{f,t}, \mathbf{N}_{f,t}\right\} \in\mathbb{C}^{P\times1}$ respectively denote the reverberant mixture, target speech and noise of $P$ channels with frequency index of $f\in\left\{1,\cdots, F\right\}$ and time index of $t\in\left\{1,\cdots,T\right\}$. Without loss of generality, the first channel is selected as the reference channel by default. $\left\{\mathbf{c}_{f}, \mathbf{r}_{f}\right\} \in\mathbb{C}^{P\times1}$ denote the relative transfer function (RTF) of the speech and that of noise, respectively. $\left\{S_{f,t}, N_{f,t}\right\}\in\mathbb{C}$ are the complex values of target speech and that of noise in the reference channel. Note that although we foucs on noise reduction in this study, it also works to directional speaker interference case, which is left as the future work.

Different from previous beamformers operated at either utterance-~{\cite{heymann2016neural}} or chunk-level~{\cite{higuchi2017online}}, we aims to estimate the framewise filter weights, which enable the real-time processing at run-time. Therefore, the beamforming output can be formulated as:
\vspace{-0.2cm}
\begin{equation}
\label{eqn2}
\widetilde{S}_{f,t} = \sum_{p=0}^{P}\left(M_{f,t}^{p}\right)^{*} X_{f,t}^{p},
\vspace{-0.3cm}
\end{equation}
where $M_{f,t}^{p}\in \mathbb{C}$ denotes the beamforming weight of the $p$th microphone. $^{*}$ stands for the conjugate operator. Note that only the noise suppression is considered and dereverberation is not addressed in this paper.
\vspace{-0.4cm}
\section{Proposed System}
\label{sec:proposed-system}
\vspace{-0.3cm}
\subsection{Forward Stream}
\vspace{-0.2cm}
In this study, the target speaker is assumed to be static within each utterance, \emph{i.e.}, the direction of arrivals (DOAs) of the target and noise remains unchanged. However, due to the highly dynamic property of speech and noise distribution, it is quite difficult to accurately separate them with spatial-only cues. To this end, we propose an embedding-and-beamforming paradigm to learn the beamforming weights from both spatial and spectral perspectives. The overall diagram of the proposed framework is shown in Fig.~{\ref{fig:architecture}}(a). It consists of three parts, namely the embedding module (EM), beamforming module (BM), and the post-processing module (PostNet). For EM, it aims to adaptively aggregate the information across the T-F spectrum and different channels and obtain the 3-D tensor where both spatial and spectral discriminative information are represented. For BM, it is employed to replace the traditional beamforming step and directly infer the filter weights and apply them to each channel. The filtered spectra are then summed together to obtain the expected speech. After the beamforming process, it may still contain the residual noise. Therefore, the PostNet is adopted to further suppress these residual noise components and improve the speech quality. In a nutshell, the whole procedure can be formulated as:
\vspace{-0.25cm}
\begin{gather}
\label{eqn3}
\widetilde{\mathbf{E}} = EMet(Cat(\mathbf{X}^{0},\cdots,\mathbf{X}^{P-1})),\\
\mathbf{\widetilde{M}} = BFNet(\mathbf{\widetilde{E}}),\\
\mathbf{\widetilde{S}}^{(1)} = \sum_{p=0}^{P}\left(\mathbf{\widetilde{M}}^{p}\right)^{\mathsf H}\mathbf{\widetilde X}^{p},\\
\mathbf{\widetilde{S}}^{(2)} = PostNet(Cat(\mathbf{\widetilde{S}}^{(1)}, \mathbf{X}^{0})),
\vspace{-0.2cm}
\end{gather}
where $\left\{EMet, BFNet, PostNet\right\}$ denote the network topology of three modules, respectively. $Cat$ refers to the concatenation operation along the channel axis. $\mathbf{\widetilde{E}}\in \mathbb{C}^{F\times T\times C}$ and $\mathbf{\widetilde{M}}\in\mathbb{C}^{F\times T\times P}$ are respectively the estimated 3-D embedding and beamforming tensors and $C$ is the embedding channel dimension. Superscripts $(1)$ and $(2)$ denote the output of the beamformer and PostNet. 
\vspace{-0.4cm}
\subsection{Embedding Module}
\label{embedding-module}
\vspace{-0.2cm}
Motivated by our preliminary works~{\cite{li2021two, li2021simultaneous}}, the convolutional ``Encoder-TCN-Decoder'' topology is employed in the EM. The encoder aims to gradually extract features with multiple downsampling operations. The decoder has the mirror structure except that all the convolution layers are replaced by the deconvolution (De) versions and recover to the original resolution. Temporal convolution networks (TCNs) serve as the bottleneck, where multiple TCMs are stacked for long-term sequence modeling and we adopt the squeezed version herein to decrease the parameter burden~{\cite{li2021two}}, \emph{i.e.}, S-TCM, as shown in Fig.~{\ref{fig:architecture}}(c).

The real and imaginary (RI) components from $P$ microphones are concatenated along the channel dimension as the network input, \emph{i.e.}, $\mathbf{X}\in\mathbb{C}^{F\times T\times 2P}$. To better capture the spatial-spectral correlation, the U$^{2}$-Encoder and U$^{2}$-Decoder are employed, which consists of multiple recalibration encoder/decoder layers (RELs/RDLs). The details are shown in Fig.~{\ref{fig:architecture}}(b)(d). Take $i$th REL/RDL as an example, it mainly consists of a 2D-(De)GLU~{\cite{dauphin2017language}}, instance normalization (IN), PReLU~{\cite{he2015delving}}, and a UNet-block with the residual connection~{\cite{qin2020u2}}. The input $\mathbf{I}_{i}$ is first encoded by the (de)convolution operation. Afterward, the UNet-block receives the encoded feature map $\mathcal{K}_{i}\left( \mathbf{I}_{i} \right)$ as the input and then further recalibrates the information distribution with a light-weight sub-UNet. For one thing, with the embedded UNet, the spectral information from multiple scales can be well grasped. For another, after consecutive downsamplings, the spatial information may get alised and the sub-UNet can effectively preserve the spatial information. The process can be given by:
\vspace{-0.4cm}
\begin{gather}
\label{eqn4}
\mathcal{K}_{i}\left(\mathbf{I}_{i}\right) = GLU\left( \mathbf{I}_{i}  \right),\\
\mathbf{O}_{i} = UNet\text{-}block\left( \mathcal{K}_{i}(\mathbf{I}_{i}) \right) + \mathcal{K}_{i}(\mathbf{I}_{i}),
\vspace{-0.2cm}
\end{gather}
\vspace{-1.0cm}
\subsection{Beamforming Module}
\label{beamforming-module}
\vspace{-0.2cm}
Having obtained the estimated 3D-embedding tensor $\mathbf{\widetilde{E}}$, the BM is leveraged to derive the framewise beamforming weights. Different from traditional beamformers, we leverage the mapping capability of networks to directly estimate the beamforming weights, which can avoid explicitly computing the spatial covariance matrix and its inversion, and thus to improve the system stability. Two types of modules are investigated as the choice herein, namely convolutional-based beamformer (C-BF) and recurrent-based beamformer (R-BF), as shown in Fig.~{\ref{fig:architecture}}(a). For the first type, a pointwise 2D-Conv is adopted to transform the channel dimension from $C$ to $2P$, which obtains the real and imaginary components of $P$-channel filters. For the second type, the layer norm (LN)~{\cite{ba2016layer}} is first adopted to normalize the embedding tensor, followed by two LSTM layers to simulate the beamforming process updated frame by frame, as shown in Fig.~{\ref{fig:architecture}}(e). Two fully-connected (FC) layers with ReLU nonlinearity activation function are used to estimate the coefficients. Note that in the previous literature~{\cite{heymann2016neural}}, the frequency dimension serves as the feature input of the LSTM. However, in this paper, the LSTM is shared by different frequency subbands, which is similar to traditional beamformers that apply to each frequency subband independently. The above process can be expressed as:
\vspace{-0.2cm}
\begin{gather}
\label{eqn5}
\mathbf{\widetilde{M}} = Conv\left(\mathbf{\widetilde{E}} \right), \text{for C-BF},\\
\mathbf{\widetilde{M}} = FC\left( LSTM\left( LayerNorm\left( \mathbf{\widetilde{E}}\right) \right) \right), \text{for R-BF},
\vspace{-0.4cm}
\end{gather}

Following filter-and-sum operation, the complex-valued filters are then applied to each channel and the filtered spectra are summed together to obtain the expected speech.
\vspace{-0.3cm}
\subsection{Post-processing Module}
\label{post-processing}
\vspace{-0.2cm}
After the beamforming stage, due to the performance limitation, some residual noise components may exist, which hinders the speech quality. To this end, a post-processing module is proposed to further suppress the remaining noise. Theoretically, any single-channel SE system can be selected. In this paper, we choose our newly proposed GaGNet as the PostNet due to its promising performance in noise reduction with low computational complexity. Due to the space limit, we may refer the readers to~{\cite{li2021glance}} for more details.
\vspace{-0.3cm}
\section{Expereimental Setup}
\label{sec:experimental-setup}
\vspace{-0.2cm}
\subsection{Dataset Preparation}
\label{dataset-preparation}
\vspace{-0.25cm}
The experiments are conducted on the DNS-Challenge dataset.{\footnote{github.com/microsoft/DNS-Challenge/tree/master/datasets}} For clean speech, the \emph{neutral clean speech} set is selected, which consists of around 562 hours by 11,350 speakers. We randomly split it into two non-overlap parts, namely for model training and evaluation. The average duration of the utterance is chunked into around 6 seconds. For the noise set, similar to~{\cite{li2021two}}, around 20,000 types of noises are randomly selected for training, whose duration is around 55 hours. Multichannel RIRs are generated with image method~{\cite{allen1979image}} based on a uniform 9-channel linear array with the spacing of 4cm. The room size is ranging from 3m-3m-2.5m to 10m-10m-3m (length-width-height). The reverberation time (RT$_{60}$) ranges from 0.05s to 0.7s. Note that the DOA difference between target speech and interference noise is at least 5$^{\circ}$ during the data generation and the distance from the source to the array is randomly selected from $\left\{0.5m, 1m, 2m, 3m\right\}$. The relative signal-to-noise ratio (SNR) between target speech and interference noise ranges $\left[-6\rm{dB}, 6\rm{dB}\right]$ with $2\rm{dB}$ interval. As a result, we create around 80,000, and 4000 mix-clean pairs for training and validation, respectively.

For model evaluation, four challenging noises are selected, namely babble, factory1, white noises from NOISEX92~{\cite{varga1993assessment}} and cafe noise from CHiME3 noise set~{\cite{barker2015third}}. Four SNRs are set, namely $\left\{ -5\rm{dB}, -2\rm{dB}, 0\rm{dB}, 2\rm{dB}  \right\}$, with 600 mix-clean pairs in each case.
\vspace{-0.8cm}
\subsection{Baselines}
\label{baselines}
\vspace{-0.2cm}
Six approaches are selected as the baselines, namely CTSNet~{\cite{li2021two}}, GaGNet~{\cite{li2021glance}}, FasNet+TAC~{\cite{luo2020end}}, MC-ConvTasNet~{\cite{zhang2020end}}, MIMO-UNet~{\cite{mimounet}}, and oracle MB-MVDR. CTSNet and GaGNet are two SE systems that achieved state-of-the-art (SOTA) performance in the monaural scenario. FasNet+TAC and MC-ConvTasNet are two time-domain based methods that exploit multiple raw-waveforms to extract the target speech. MIMO-UNet ranked first in the Far-field Multi-Channel Speech Enhancement Challenge for Video Conferencing. For a fair comparison, all the non-causal settings are replaced by the causal versions. Besides, the scale-invariant SNR (SI-SNR) loss in FasNet+TAC and MC-ConvTasNet is replaced by classical SNR loss to mitigate the free change of the magnitude level~{\cite{kinoshita2020improving}}. For MB-MVDR, we utilize the oracle ideal ratio mask (IRM) to calculate the spatial covariance matrix and then derive MVDR weights for beamforming.{\footnote{https://pypi.org/project/beamformers}
\vspace{-0.3cm}
\subsection{Implementation Details}
\label{implementation-details}
\vspace{-0.1cm}
\subsubsection{Model Details}
\vspace{-0.2cm}
In the EM, the kernel size of 2D-(De)GLU is set as  $\left(2,3\right)$ with stride being $\left(1,2\right)$ in the time and frequency axes. For each UNet-block, the kernel and stride size are $\left(1,3\right)$ and $\left(1,2\right)$, respectively. Let us define the number of (de)encoding layers within the UNet-block as $Q$, then $Q=\left\{4,3,2,1,0\right\}$ for U$^{2}$-Encoder and $Q=\left\{1,2,3,4,0\right\}$ for U$^{2}$-Decoder, respectively, where $0$ means that no UNet-block is used. The number of channels in both encoder and decoder remains 64 by default. For bottleneck sequence modeling, 3 S-TCNs are stacked, each of which consists of 6 S-TCMs with kernel size and dilation rate being 5 and $\left\{1,2,4,8,16,32\right\}$. When R-BF is switched for beamforming, two uni-directional LSTMs are utilized with 64 hidden nodes. For C-BF, the kernel size is set to $(1,1)$.
\vspace{-0.4cm}
\subsubsection{Training Details}
\vspace{-0.2cm}
All the utterances are sampled at 16kHz. The 20ms Hanning window is utilized with 50\% overlap between adjacent frames. 320-point FFT is utilized, leading to 161-D features, \emph{i.e.}, $F$=161. Most recently, the efficacy of power-compression is investigated for single-channel speech enhancement~{\cite{li2021glance}} and dereverberation tasks~{\cite{li2021importance}}. Here, we adopt it for both the input and target of each channel, \emph{i.e.}, $|\mathbf{X}^{p}|^{0.5}e^{j\theta_{\mathbf{X}^{p}}}$, $|\mathbf{S}^{p}|^{0.5}e^{j\theta_{\mathbf{S}^{p}}}, p\in\left\{1\cdots,P \right\}$. The rationale is that we only compress the magnitude and leave the phase unchanged. Therefore, spatial information can still be well preserved. MMSE with magnitude constraint is adopted as the loss function for training~{\cite{li2021two, li2021glance}}. All the models are trained with Adam optimizer~{\cite{kingma2014adam}} and the learning rate is initialized at 5e-4 and will be halved if the loss does not decrease for consecutive two epochs. The batch size is 8 and the number of epochs is 60.
\vspace{-0.4cm}

\renewcommand\arraystretch{0.65}
\begin{table*}[t]
	\caption{Ablation study on the proposed EaBNet. The values are specified with PESQ/ESTOI(\%)/SDR(dB) format. \textbf{BOLD} indicates the best score in each case. ``Avg.'' denotes the average value among different SNRs in the test set.}
	\tiny
	\setlength{\tabcolsep}{3pt}
	\centering
	\resizebox{0.890\textwidth}{!}{
		\begin{tabular}{c|c|ccccc|ccccc}
			\hline
			\multirow{2}*{System} &\multirow{2}*{ID} &UNet-
			&\multirow{2}*{MO} &BF  &\multirow{2}*{Compression} &Para.  &\multirow{2}*{-5\rm{dB}} &\multirow{2}*{-2\rm{dB}} &\multirow{2}*{0\rm{dB}} &\multirow{2}*{2\rm{dB}} &\multirow{2}*{Avg.}\\
			 & &block & &Type & &(M) & & & & &\\
			\cline{1-12}
			\multirow{5}*{EaBNet} &1 &\XSolidBrush &\Checkmark &R-BF &\Checkmark &\textbf{2.19} &3.16/78.95/13.83 &3.34/82.36/15.45 &3.49/86.18/16.90 &3.59/87.63/17.55 &3.40/83.78/15.93\\
			 &2 &\Checkmark &\XSolidBrush &\XSolidBrush &\Checkmark &2.77 &3.10/77.22/12.26 &3.28/80.80/13.74 &3.44/84.65/15.09 &3.54/86.28/15.94 &3.34/82.24/14.26\\
		    &3 &\Checkmark &\Checkmark &C-BF &\Checkmark &2.77 &3.20/79.67/13.40 &3.38/82.97/15.05 &3.54/86.68/16.50 &3.63/87.89/17.12 &3.44/84.30/15.52\\
			 &4 &\Checkmark &\Checkmark &R-BF &\XSolidBrush &2.84 &2.93/76.66/\textbf{14.73} &3.12/80.88/\textbf{16.53} &3.29/84.83/\textbf{18.13} &3.39/86.21/\textbf{18.61} &3.18/82.15/\textbf{17.00} \\
			 &5 &\Checkmark &\Checkmark &R-BF &\Checkmark &2.84 &\textbf{3.30}/\textbf{81.75}/14.68 &\textbf{3.47}/\textbf{84.66}/16.16 &\textbf{3.61}/\textbf{88.04}/17.64 &\textbf{3.70}/\textbf{89.19}/18.38 &\textbf{3.52}/\textbf{85.91}/16.72\\
			\hline
	\end{tabular}}
	\label{tbl:ablation-study}
	\vspace{-0.6cm}
\end{table*}
\renewcommand\arraystretch{1.10}
\begin{table*}[t]
	\caption{Results comparison with advanced baselines. ``Cau.'' denotes whether the system is causal implementation.}
	\tiny
	\setlength{\tabcolsep}{3pt}
	\centering
	\resizebox{0.925\textwidth}{!}{
		\begin{tabular}{c|ccccc|ccccc}
			\hline
			\multirow{2}*{Systems} &\multirow{2}*{Domain} &Para. &MACs
			&\multirow{2}*{RTF} &\multirow{2}*{Cau.}  &\multirow{2}*{-5\rm{dB}} &\multirow{2}*{-2\rm{dB}} &\multirow{2}*{0\rm{dB}} &\multirow{2}*{2\rm{dB}} &\multirow{2}*{Avg.}\\
			& &(M) &(G/s) & & & & & & & \\
			\cline{1-11}
			Noisy &- &- &- &- &- &1.45/29.62/-4.89 &1.62/37.71/-1.93 &1.74/41.85/0.05 &1.87/49.42/2.05 &1.67/39.65/-1.18 \\
			CTSNet &T-F &4.35 &5.57 &0.37  &\Checkmark &1.87/40.35/2.32 &2.14/50.87/5.63 &2.34/58.27/7.64 &2.50/63.64/9.03 &2.21/53.28/6.15\\
			GaGNet &T-F &5.94 &\textbf{1.63} &0.19 &\Checkmark &1.91/42.02/3.01 &2.22/52.50/5.99 &2.42/59.99/7.90 &2.58/65.09/9.19 &2.28/54.90/6.52\\
			FasNet+TAC &T &3.82 &7.56 &0.67 &\Checkmark &2.40/63.39/11.43 &2.62/68.99/13.44 &2.77/73.89/14.88 &2.88/76.50/15.54 &2.67/70.69/13.82\\
			MC-ConvTasnet &T &6.56 &5.28 &0.43 &\Checkmark &2.21/59.57/9.90 &2.44/65.20/11.61 &2.72/72.16/13.46 &2.82/74.52/14.02 &2.55/67.86/12.25\\
			MIMO-UNet &T-F &\textbf{1.97} &4.09 &\textbf{0.16} &\Checkmark &2.39/60.93/8.96 &2.61/66.99/11.17 &2.75/71.29/12.49 &2.85/73.71/13.20 &2.65/68.23/11.45\\
			MB-MVDR(oracle) &T-F &- &- &- &\XSolidBrush &2.88/78.59/12.06 &3.04/82.45/13.90 &3.18/85.82/15.17 &3.29/87.43/15.90 &3.10/83.57/14.26\\
			\hline
			EaBNet$^{\ast}$ &T-F &2.91 &8.46 &0.80 &\Checkmark &3.24/80.16/13.60 &3.41/83.49/15.15 &3.56/86.91/16.51 &3.65/88.11/17.15 &3.46/84.67/15.60\\
			EaBNet &T-F &2.84 &7.38 &0.59 &\Checkmark &3.30/81.75/14.68 &3.47/84.66/16.16 &3.61/88.04/17.64 &3.70/89.19/18.38 &3.52/85.91/16.72\\
			EaBNet+PostNet &T-F &8.78 &9.04 &0.83 &\Checkmark &\textbf{3.44}/\textbf{83.33}/\textbf{15.13} &\textbf{3.58}/\textbf{85.86}/\textbf{16.58} &\textbf{3.71}/\textbf{89.04}/\textbf{18.05} &\textbf{3.79}/\textbf{90.03}/\textbf{18.72} &\textbf{3.63}/\textbf{87.06}/\textbf{17.12}\\
			\hline
	\end{tabular}}
	\label{tbl:results-comparison}
	\vspace{-0.5cm}
\end{table*}
\section{RESULTS AND ANALYSIS}
\label{sec:results-and-analysis}
\vspace{-0.2cm}
\subsection{Ablation Study}
\label{ablation-study}
\vspace{-0.2cm}
We conduct the ablation study on EaBNet in terms of whether to use UNet-block, whether to output multiple filter weights (MO), the type of BF, and whether the magnitude compression is adopted, as shown in Table~{\ref{tbl:ablation-study}}. Perceptual evaluation of speech quality (PESQ)~{\cite{rix2001perceptual}}, extended short-time objective intelligibility (ESTOI)~{\cite{jensen2016algorithm}}, and signal-distortion ratio (SDR)~{\cite{vincent2007first}} are adopted as evaluation metrics. Several observations can be made. 1) Going from ID-1 to ID-5, constant improvements are made for all cases, which show that the introduction of the UNet-block can well preserve spectral and spatial information and lead to better beamforming results. 2) When we directly estimate the complex-valued mask for the reference channel without explicit beamforming process, as shown from ID-5 to ID-2, consistent performance degradations in three metrics are observed, which emphasize the significance of the beamforming operation in multi-channel speech enhancement systems. 3) We compare the performance between different BF types, as shown in ID-3 and ID-5, one can find that R-BF yields relatively better performance over C-BF. This is because, in R-BF, the LSTM is leveraged to update the state frame by frame, which leads to better beamforming weights estimation when the spatial information is not accurate enough for separation. 4) Compared with ID-4, when the magnitude compression is employed, considerable improvements in PESQ and ESTOI are achieved while mild degradation in SDR is observed. This is because compression operation decreases the dynamic range of spectrum distribution and highlights the priority of low-energy regions~{\cite{li2021glance,li2021importance}}. As the result, more residual noise can be suppressed and improve the speech quality. Meanwhile, with the nonlinear compression operation, the linear separability between different sources in the space may be destroyed. Therefore, it may cause more target distortion, which partly explains the degradation in the SDR.
\vspace{-0.4cm}
\subsection{Results Comparison with Advanced Baselines} 
\label{results-comparison}
\vspace{-0.2cm}
The best configuration of EaBNet in Table~{\ref{tbl:ablation-study}} is chosen to compare with other baselines, whose results are presented in Table~{\ref{tbl:results-comparison}}. To emphasize the effectiveness of the learned embedding in spectral-spatial information representation, we also set the reference dubbed EaBNet$^{\ast}$, where we output the complex-valued masks \emph{w.r.t.} speech and noise at the end of EB and then the spatial covariance matrices are calculated and concatenated as the input of R-BF~{\cite{xu2021generalized}}. For a fair comparison, the hidden nodes of the LSTM remain the same.

From the table, several observations can be obtained. First, compared with the single-channel case, when more channels are available, considerable improvements for three metrics can be achieved for all the multichannel based models. This indicates that the utilization of spatial information can facilitate the separation of different sources. Second, the proposed system outperforms the previous baselines by a large margin. For example, going from MIMO-UNet to EaBNet, average 0.87, 17.68\%, and 5.67dB improvements are achieved in terms of PESQ, ESTOI, and SDR, respectively. It fully demonstrates the superiority of our system in speech recovery. Besides, we observe that the proposed system also surpasses the MB-MVDR with oracle IRM consistently, which reveals the feasibility of end-to-end framewise beamformers with DNNs over previous tandem-style schemes~{\cite{heymann2016neural, erdogan2016improved, heymann2015blstm}}. Third, it is interesting to find that compared with EaBNet$^{\ast}$, when the embedding tensor is abstractly represented rather than follow the traditional signal-theory to calculate the spatial covariance matrices \emph{w.r.t.} speech and noise, even better performance can be made, which inspires us to rethink the role of signal-theory in end-to-end neural beamformers. We can explain the phenomenon from several perspectives. For one thing, the spatial covariance matrix is usually sparse and is often redundant and not necessary to represent the spectral-temporal information. Meanwhile, it also tends to be less robust toward the real scenarios than the compact embedding. For another, compared with explicit second-order covariance matrix calculation, the implicit embedding is learned directly from the training data, it can thus potentially learn higher-order spatial statistics. Forth, when the PostNet is adopted, further metric improvements can be achieved, which illustrates the necessity of post-processing in noise suppression and speech recovery.

We also provide the model size, the number of multiply-accumulate operations (MACs) per second, and real-time factor (RTF), as shown in Table~{\ref{tbl:results-comparison}}. RTF is evaluated on an Intel Core(TM) i5-4300 CPU clocked at 1.90GHz. One can find that EaBNet has an overall light-weight model size (2.84M) than other baselines and the RTF is 0.59, which meets the real-time processing criterion. Despite more parameters and higher RTF are induced when the PostNet is added, we can decrease the overall burden by choosing more decent post-processing algorithms with less computational complexity.
\vspace{-0.3cm}
\section{CONCLUSIONS}
\label{sec:conclusion}
\vspace{-0.2cm}
In this paper, we propose a generalized causal framework called EaBNet, which enables the framewise neural beamforming for multichannel speech enhancement. Two modules are designed accordingly, namely EM and BM. In the EM, we directly generate the 3-D embedding tensor which contains both spectral-spatial discriminative information. In the BM, a network is directly utilized to output the filter weights. A post-processing module is also introduced to further suppress the residual noise and facilitate speech recovery. The experiments show that the proposed system yields state-of-the-art performance over previous baselines by a large margin.



\vfill\pagebreak

\bibliographystyle{IEEEbib}

\end{document}